\documentclass{iaus}
\usepackage{graphicx}

\newcommand{\FeI}{\mbox{Fe\,{\sc i}}}
\newcommand{\Halpha}{\mbox{H\hspace{0.1ex}$\alpha$}}

\title[flare induced penumbral formation] 
{ Flare induced penumbra formation in the sunspot of NOAA 10838}

\author[Sreejith P. \& Sankarasubramanian K.]   
{Sreejith Padinhatteeri$^{1,2}$
 \and Sankarasubramanian K.$^1$}

\affiliation{$^1$Space Astronomy Group, ISRO Satellite Centre, Bangalore,\\ India - 560017 \\ email:
{sreejith.p@gmail.com, sankark@isac.gov.in} \\[\affilskip]
$^2$Dept. of Physics, University of Calicut, Kerala, India. }

\pubyear{2010}
\volume{273}  
\pagerange{1--4}
\jname{Title of your IAU Symposium}
\editors{A.C. Editor, B.D. Editor \& C.E. Editor, eds.}
\begin{document}

\maketitle

\begin{abstract}

We have observed formation of penumbrae on a pore in the active region NOAA10838
using Dunn Solar Telescope at NSO,Sunpot,USA. Simultaneous observations using
different instruments (DLSP,UBF,Gband and CaK) provide us with vector	magnetic
field at photosphere, intensity images and Doppler velocity at different heights
from photosphere to chromosphere. Results from our analysis of this particular
data-set suggests that penumbrae are formed as a result of  relaxation of
magnetic field due to a flare happening at the same time.  Images in \Halpha\
show the flare (C 2.9 as per GOES) and vector magnetic fields show a
re-orientation and reduction in the global $\alpha$ value (a measure of twist).
We feel such relaxation of loop structures due to reconnections or flare could
be one of the way by which field lines fall back to the photosphere to form
penumbrae.

\end{abstract}

\firstsection 
\section{Introduction}

Sunspots are the manifestation of strong magnetic fields that emerge in the
solar photosphere (\cite{1964suns.book.....B}, \cite{2003A&ARv..11..153S}).
Although sunspots are stable configurations when compared to the dynamical time
scales of other features on the Sun, the observed umbral and penumbral
fine-structures are very dynamic and subjected to constant change and
transformation on small spatial scales. Our understanding of these processes and
the nature of the fine structures improved significantly in the last decade
(\cite{2008ApJ...672..684R}, \cite{2008ApJ...686.1454B}), but we still lack
detailed knowledge about the key process of penumbral formation and decay.
Observations of this process are very rare, most prominent among them are by
\cite{1998ApJ...507..454L}, \cite{yang03} and \cite{schlichenmaier2010}. All of
them observe the formation of a penumbra happening in few hours time. In all the
cases they suggest the onset of penumbral formation is due the flux emergence.
\cite{1998ApJ...507..454L} suggests critical flux limit of $1-1.5\times10^{20}$ Mx for
initiation of penumbra. Here in this paper we discuss multi-wavelength
observations of the formation of penumbrae in the active region NOAA 10838
which was carried out on 2005 December 22.  Coincidentally, a C-2.9 class flare
was also observed in the same active region. These observations corresponding to
different heights in the solar atmosphere suggest that the penumbral formation
in this particular case is related to the flare at chromospheric and higher
layers.
\begin{figure}
\begin{center}
 \includegraphics[width=4.2in]{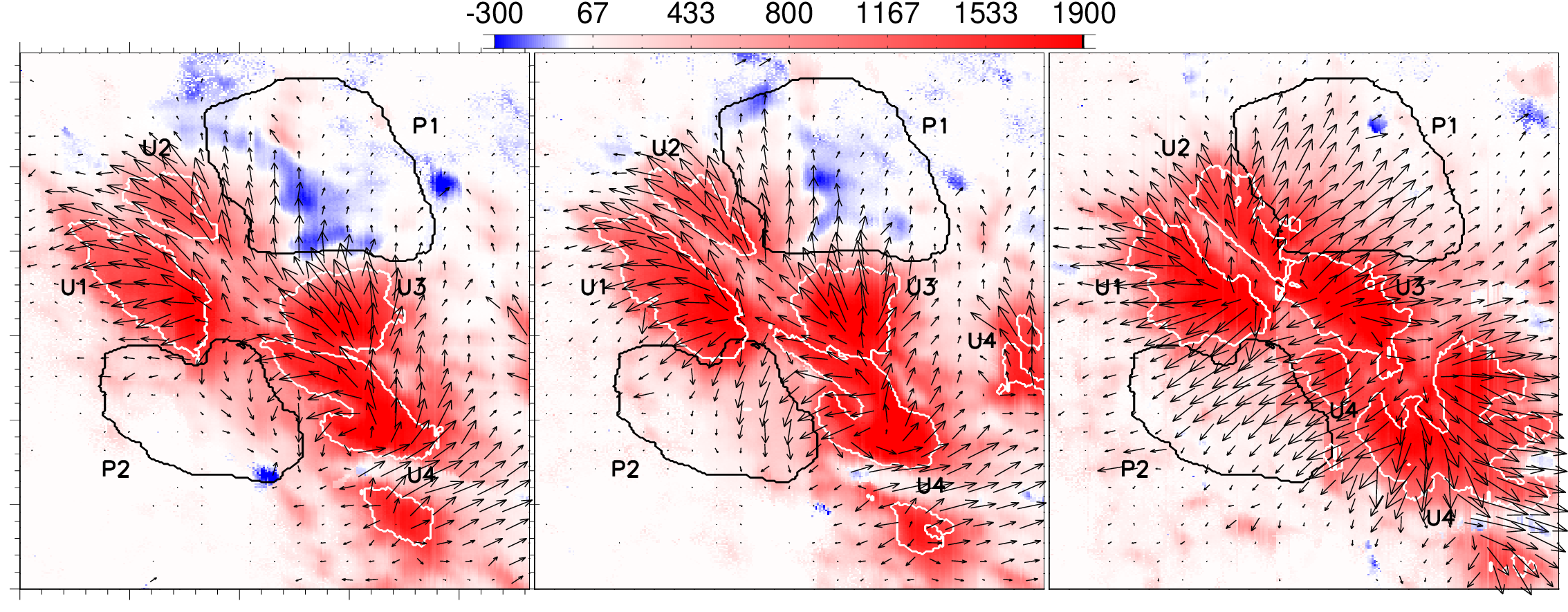}
 \includegraphics[width=4.2in]{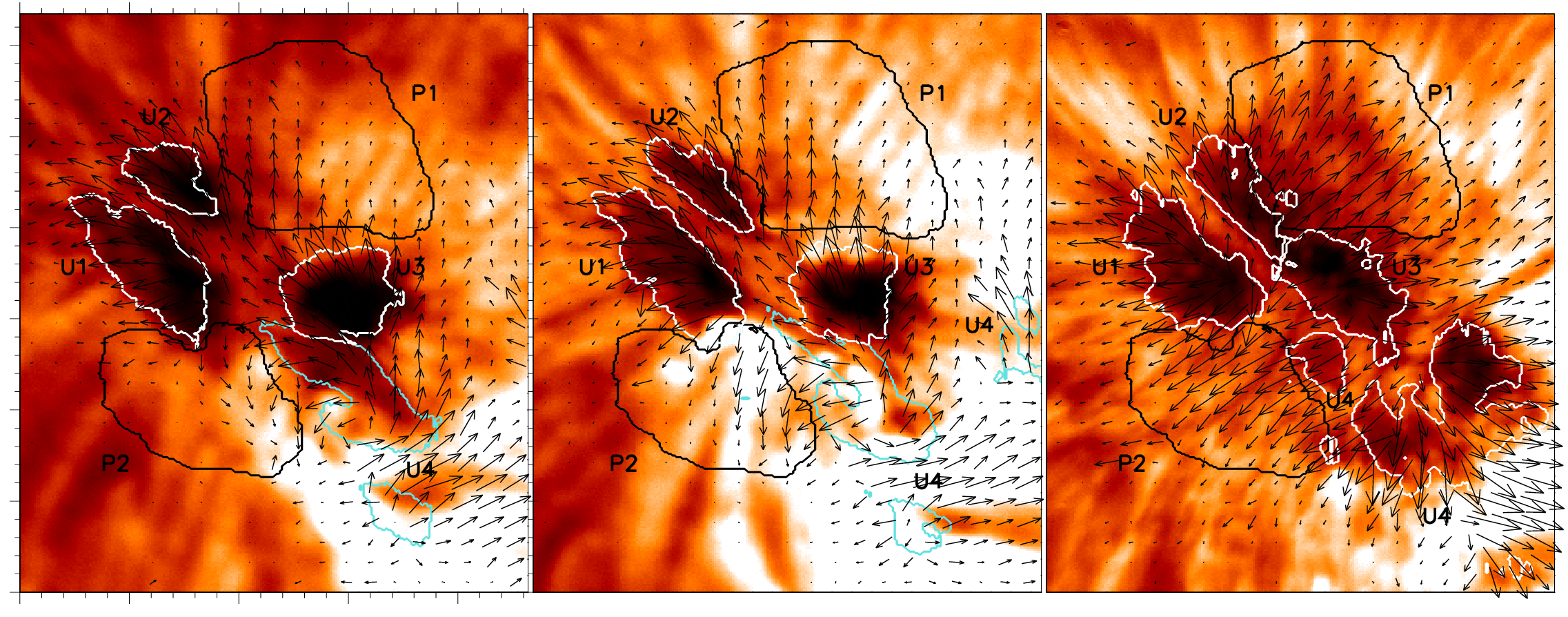}
 \includegraphics[width=4.2in]{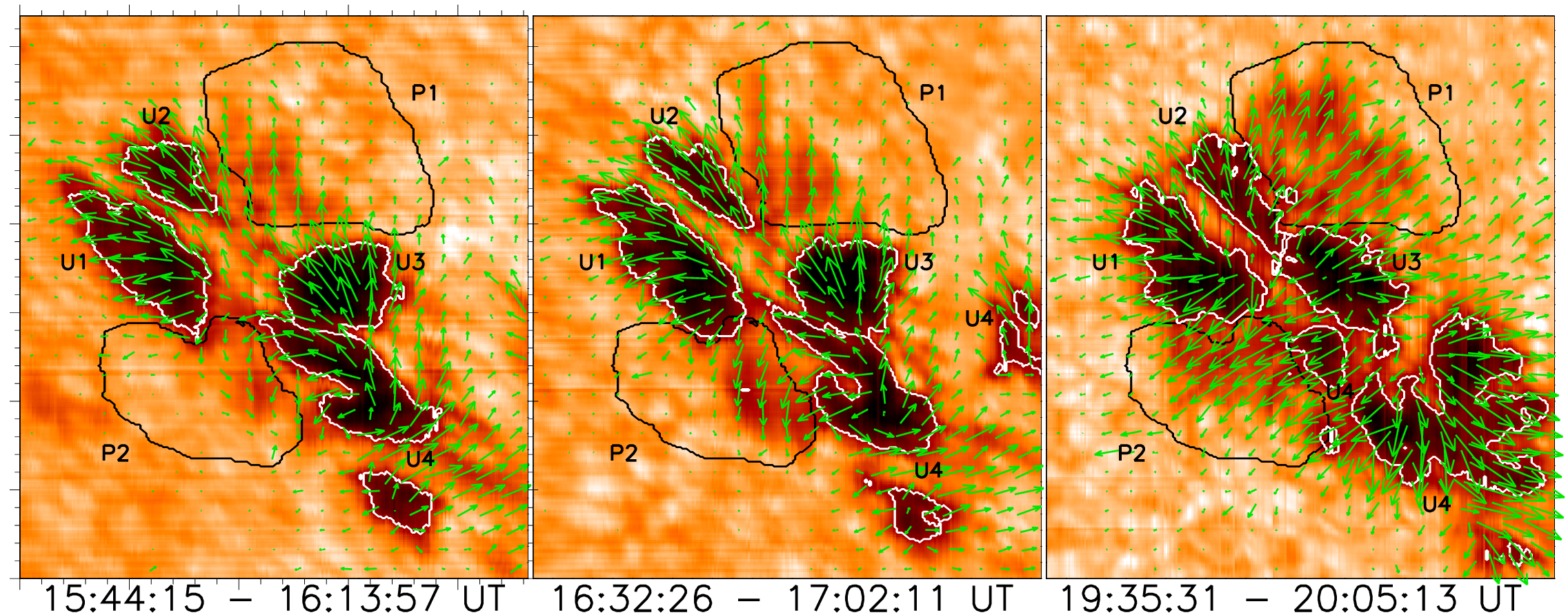}
 \caption{Figures showing Vector magnetic field evolution during the penumbral
 formation. Horizontal field vectors (arrows) are overplotted on vertical magnetic
 field (top row; with colour bar on the top showing the $B_z$ values),
 \Halpha\ image (middle row) and continuum image (bottom row). Raster scan time is mentioned
 at the bottom of each column. Dark contours mark the area where penumbra forms,
 and white contours mark different umbrae.}
   \label{fig1}
\end{center}
\end{figure}
\section{Observations and Data Analysis}
The formation of the penumbra was observed on the leading sunspot of the active
region NOAA 10838 on 2005 December 22 at the Dunn Solar Telescope in Sunspot,NM,
USA. The observed sunspot was situated at ${\mu} = cos{\theta} = 0.77$~(N15.5
E35.4) on the solar disk. The atmospheric seeing was moderate during the
observation. The adaptive optics system at the DST (\cite{2004SPIE.5171..179R})
was operated and the atmospheric seeing corrected beam was fed to a set of
back-end instruments. The main back-end instruments used were: Diffraction
Limited Spectro-Polarimeter (DLSP; \cite{2006ASPC..358..201S}) and Universal
Birefringent Filter (UBF; \cite{ubfreport1975}). DLSP was used to obtain stokes
profiles of two magnetically sensitive lines, \FeI\ $\lambda\ 6301.5$ \AA\ and
$6302.5$ \AA.\ The stokes profiles were inverted with an assumption of
Milne-Eddington atmosphere using the HAO inversion code. The magnetic field
vector components were transformed from observer's to heliographic
co-ordinates using simple spherical trigonometric transformations (see
 \cite{smart}).  The UBF is a tunable Lyot filter with a pass-band that varies
between 120 to 250 m\AA\ in the visible wavelength of 5000 to 7000 \AA. The UBF
was tuned in to two different spectral lines namely, \FeI\ $\lambda\ 5434$ and
\Halpha\ $\lambda\ 6562$.  We have also used the archived data from EIT and MDI
onboard SOHO satellite, during the same observation time, to study the loop
dynamics during the flare.  EIT and MDI data were calibrated and matched each
other using standard codes available in the SSW software package. Magnetic field
contours derived from MDI were overplotted on EIT images. To bring up the loop
structure, a time averaged EIT image was subtracted from each frame and the
intensity level contours were overplotted.

\section{Results and conclusion}

We have observed the formation of penumbrae, and a flare happening
simultaneously. Preliminary results suggests a relation between the two.  Top
row of the Figure~\ref{fig1} shows the magnetic filed orientations at three
raster scan maps. The colour code is used for vertical field strength and the
arrows represent horizontal field.  Dark Solid contours mark the ROI, selected
manually, where penumbral formation happens, and are marked as {\bf P1} and {\bf
P2}. Four umbral areas are selected using the intensity values (less than 0.7
$I_{QS}$ where $I_{QS}$ is the quite sun mean intensity), and are marked in
white contours. U1, U2 and U3 are the bottom left, top left, and top right
umbrae respectively. The bottom right umbra undergoes lot of changes, in shape
and area, and many other small pores come and join. Hence all the umbral area
other than U1, U2 and U3 are marked as U4. Apart from the bottom right umbra
seen in first column, small pores which later join this spot are all grouped
under U4. In the second row the horizontal field vector is overplotted on
\Halpha\ line core images obtained using UBF at similar time of the respective
raster scan. The flare can be seen as high intensity values in \Halpha\ images.
In the third row horizontal field vector is overplotted on continuum raster scan
image.  The formation of penumbra is clearly visible in the continuum images.
It is clear from the middle and bottom row images that the horizontal vectors
match quite well with the magnetic structures like penumbra seen in continuum
images and the super penumbral structures in \Halpha\ line core images. This
suggest that the performed analyses, like inversions, azimuth ambiguity
corrections and coordinate transformation from observers to heliographic
co-ordinate were proper.

\begin{figure}
\begin{center}
 \includegraphics[width=5.05in]{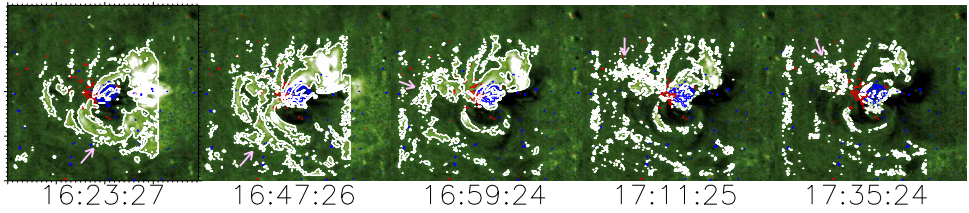}
 \caption{ EIT images with white contours highlighting the bright loops. Red and
 blue contours represent positive and negative magnetic field, as derived from
 MDI, respectively. The arrow in each image show the loop that untwist as the
 flare happens.}
   \label{fig2}
\end{center}
\end{figure}
The arrows, representing the horizontal magnetic field, show a twisted field
orientation in first two maps (first and second column of Figure~\ref{fig1}),
and as the flare happens, the field lines untwist to a more potential like
nature as seen in third column. The global alpha value, {\boldmath$\alpha_{g}$},
which is a measure of magnetic field twist was estimated (see for e.g. \cite{
sanjiv09}). The {\boldmath$\alpha_{g}$} of this sunspot reduces from
$1.77\times10^{-7}/m$ to $-0.43\times10^{-7}/m$ as the penumbrae forms.  The
reduction in the {\boldmath$\alpha_{g}$} shows the relaxation of magnetic field
configuration at the photosphere, due to the flare. The flare was of C-2.9 class
as measured by GOES.  Flare started at around 16:32 UT and peaked at around
16:50 UT.

The total longitudinal magnetic flux ($\sum {f B_z A}$, where $f$ is the
fill fraction, $A$ is the area and $B_z$ is the longitudinal magnetic field
intensity) of
this spot is of the order  $10^{21}$ Mx. It's little higher than the critical
limit of $10^{20}$Mx needed for penumbral formation as suggested by
\cite{1998ApJ...507..454L}. There is an increase in the flux majorly due to
similar polarity pores joining the main spot in U4. The flux values of four
umbral region and the penumbral forming region are tabulated in
Table~\ref{tbl1}. Apart from the flux values one also have to consider the 
twisted configuration of the magnetic fields.  We believe that, in this case, 
the flare which transformed the twisted
field structure to a more potential nature might have triggered the formation of
penumbra.
\begin{table}
\caption{\label{tbl1}Magnetic Flux $\Phi$ in units of $10^{20}$Mx.  Errors
to Flux values in umbra are less than $0.05\%$ and in penumbra are less than
$0.12\%$}
\begin{center}
\begin{tabular}{ccccccc}
\hline
     &U1 & U2 & U3 & U4 & P1 & P2 \\
\hline

    Map-1 & 1.67 & 0.50 & 1.45 & 2.44 & 0.37 & 0.89 \\
    Map-2 & 1.96 & 0.61 & 1.55 & 2.65 & 0.45 & 0.99 \\
    Map-3 & 2.06 & 1.02 & 1.89 & 3.76 & 1.58 & 0.74 \\
\hline
\end{tabular}
\end{center}
\end{table}

Figure~\ref{fig2} gives the five images taken using EIT, with $\approx$ 12
minutes cadence. The white contours shows the loops, and the blue and red
contours represent the negative and positive magnetic field strength, derived
from MDI, respectively. The images were taken close to the peak of the flare.
The images show untwisting of large loops, in the same direction as seen in the
photosphere vector magnetic field configuration. Similar untwisting is also seen
in \Halpha\ time sequence movie. It suggest that the untwisting of magnetic
field and the subsequent relaxation happens all the way from corona to
photosphere.

We conclude that, observation of this particular event on this sunspot suggest a
relation between flare and penumbral formation. Apart from the coincidence of
both events, the observation also confirms the magnetic field re-orientation,
all the way from corona to photosphere. Hence, we speculate that the penumbral
formation may not be simply a photospheric phenomena but a result of global
reconfiguration of magnetic field due to activities like a flare.\\

\noindent
{\bf Acknowledgements:} SOHO is a project of international cooperation between
NASA and ESA. The NSO is operated by the AURA under a cooperative agreement with
NSF.


\begin{thebibliography}{}

\bibitem[{{{Beckers} \textit{et~al.}} (1975)}]{ubfreport1975} {Beckers}, J.M.,
    {Dickson}, L., {Joyce}, R.S., 1975, AFCRL Report No. AFCRL-TR-75-0090,A ir
    Force Cambridge Research Laboratory, Massachusetts.

\bibitem[{{Bray} \& {Loughhead} (1964)}]{1964suns.book.....B} {Bray}, R.~J.
    {Loughhead}, R.~E. 1964, {Sunspots} (The International Astrophysics Series,
    London: Chapman {} Hall, 1964)

\bibitem[{{Brummell} \textit{et~al.} (2008)}]{2008ApJ...686.1454B} {Brummell}, N.~H.,
    {Tobias}, S.~M., {Thomas}, J.~H., {Weiss}, N.~O. 2008, \textit{ApJ},686,
    1454

\bibitem[{{Leka} \& {Skumanich} (1998)}]{1998ApJ...507..454L}
    {Leka}, K.~D. {Skumanich}, A. 1998, \textit{ApJ}, 507, 454

\bibitem[{{Rimmele} (2008)}]{2008ApJ...672..684R} {Rimmele}, T. 2008,
    \textit{ApJ}, 672, 684

\bibitem[{{Rimmele} \textit{et~al.} (2004)}]{2004SPIE.5171..179R} {Rimmele},
    T.~R., {Richards}, K., {Hegwer}, S., {et~al.} 2004, in SPIE Conf. Ser., eds.
    S.~{Fineschi} \& M.~A. {Gummin}, \textit{SPIE Conf. Ser.}, 5171, 179

\bibitem[{{Sankarasubramanian} \textit{et~al.} (2006)}]{2006ASPC..358..201S}
    {Sankarasubramanian}, K., {Lites}, B., {Gullixson}, C., \textit{et~al.}
    2006, in ASP Conf. Ser., eds.  R.~{Casini} \& B.~W. {Lites}, \textit{ASP
    Conf. Ser.}, 358, 201

\bibitem[{{Schlichenmaier} \textit{et~al.} (2010)}] {schlichenmaier2010}
    {Schlichenmaier}, R.,{Rezaei}, R., {Gonz\'{a}lez}, N.~B., {Waldmann}, T.~A.
    2010, \textit{A\&A}, 512, L1

\bibitem[{{Smart} \& {Green} (1977)}]{smart}{Smart}, W.~M., {Green}, R.~M.
    1977, {Textbook on Spherical Astronomy}, (Cambidge University Press,
    Cambidge, 1977)

\bibitem[{{Solanki} (2003)}]{2003A&ARv..11..153S} {Solanki}, S.~K.
    2003,\textit{A\&AR}, 11, 153

\bibitem[{{Tiwari} \textit{et~al.} (2009)}]{sanjiv09} {Tiwari}, S.~K., {Venkatakrishnan},
    P., {Gosain}, S., \& {Joshi}, J. 2009, \textit{ApJ}, 700, 199

\bibitem[{{Yang},\textit{et~al.} (2003)}]{yang03} {Yang}, G., {Xu}, Y., {Wang}, H.,
    {Denker}, C. 2003, \textit{ApJ}, 597,1190


\end{thebibliography}
\end{document}